\def\sphinxdocclass{article}
\documentclass[a4paper,10pt,english]{sphinxhowto}
\usepackage[utf8]{inputenc}
\DeclareUnicodeCharacter{00A0}{\nobreakspace}
\usepackage[T1]{fontenc}
\usepackage{babel}
\usepackage{times}
\usepackage[Bjarne]{fncychap}
\usepackage{longtable}
\usepackage{sphinx}

\title{TiA 1.0}
\date{March 23, 2011}
\release{}
\author{Christian Breitwieser, Christoph Eibel}
\newcommand{\sphinxlogo}{}
\renewcommand{\releasename}{Release}
\makeindex

\makeatletter
\def\PYG@reset{\let\PYG@it=\relax \let\PYG@bf=\relax%
    \let\PYG@ul=\relax \let\PYG@tc=\relax%
    \let\PYG@bc=\relax \let\PYG@ff=\relax}
\def\PYG@tok#1{\csname PYG@tok@#1\endcsname}
\def\PYG@toks#1+{\ifx\relax#1\empty\else%
    \PYG@tok{#1}\expandafter\PYG@toks\fi}
\def\PYG@do#1{\PYG@bc{\PYG@tc{\PYG@ul{%
    \PYG@it{\PYG@bf{\PYG@ff{#1}}}}}}}
\def\PYG#1#2{\PYG@reset\PYG@toks#1+\relax+\PYG@do{#2}}

\def\PYG@tok@gd{\def\PYG@tc##1{\textcolor[rgb]{0.63,0.00,0.00}{##1}}}
\def\PYG@tok@gu{\let\PYG@bf=\textbf\def\PYG@tc##1{\textcolor[rgb]{0.50,0.00,0.50}{##1}}}
\def\PYG@tok@gt{\def\PYG@tc##1{\textcolor[rgb]{0.00,0.25,0.82}{##1}}}
\def\PYG@tok@gs{\let\PYG@bf=\textbf}
\def\PYG@tok@gr{\def\PYG@tc##1{\textcolor[rgb]{1.00,0.00,0.00}{##1}}}
\def\PYG@tok@cm{\let\PYG@it=\textit\def\PYG@tc##1{\textcolor[rgb]{0.25,0.50,0.56}{##1}}}
\def\PYG@tok@vg{\def\PYG@tc##1{\textcolor[rgb]{0.73,0.38,0.84}{##1}}}
\def\PYG@tok@m{\def\PYG@tc##1{\textcolor[rgb]{0.13,0.50,0.31}{##1}}}
\def\PYG@tok@mh{\def\PYG@tc##1{\textcolor[rgb]{0.13,0.50,0.31}{##1}}}
\def\PYG@tok@cs{\def\PYG@tc##1{\textcolor[rgb]{0.25,0.50,0.56}{##1}}\def\PYG@bc##1{\colorbox[rgb]{1.00,0.94,0.94}{##1}}}
\def\PYG@tok@ge{\let\PYG@it=\textit}
\def\PYG@tok@vc{\def\PYG@tc##1{\textcolor[rgb]{0.73,0.38,0.84}{##1}}}
\def\PYG@tok@il{\def\PYG@tc##1{\textcolor[rgb]{0.13,0.50,0.31}{##1}}}
\def\PYG@tok@go{\def\PYG@tc##1{\textcolor[rgb]{0.19,0.19,0.19}{##1}}}
\def\PYG@tok@cp{\def\PYG@tc##1{\textcolor[rgb]{0.00,0.44,0.13}{##1}}}
\def\PYG@tok@gi{\def\PYG@tc##1{\textcolor[rgb]{0.00,0.63,0.00}{##1}}}
\def\PYG@tok@gh{\let\PYG@bf=\textbf\def\PYG@tc##1{\textcolor[rgb]{0.00,0.00,0.50}{##1}}}
\def\PYG@tok@ni{\let\PYG@bf=\textbf\def\PYG@tc##1{\textcolor[rgb]{0.84,0.33,0.22}{##1}}}
\def\PYG@tok@nl{\let\PYG@bf=\textbf\def\PYG@tc##1{\textcolor[rgb]{0.00,0.13,0.44}{##1}}}
\def\PYG@tok@nn{\let\PYG@bf=\textbf\def\PYG@tc##1{\textcolor[rgb]{0.05,0.52,0.71}{##1}}}
\def\PYG@tok@no{\def\PYG@tc##1{\textcolor[rgb]{0.38,0.68,0.84}{##1}}}
\def\PYG@tok@na{\def\PYG@tc##1{\textcolor[rgb]{0.25,0.44,0.63}{##1}}}
\def\PYG@tok@nb{\def\PYG@tc##1{\textcolor[rgb]{0.00,0.44,0.13}{##1}}}
\def\PYG@tok@nc{\let\PYG@bf=\textbf\def\PYG@tc##1{\textcolor[rgb]{0.05,0.52,0.71}{##1}}}
\def\PYG@tok@nd{\let\PYG@bf=\textbf\def\PYG@tc##1{\textcolor[rgb]{0.33,0.33,0.33}{##1}}}
\def\PYG@tok@ne{\def\PYG@tc##1{\textcolor[rgb]{0.00,0.44,0.13}{##1}}}
\def\PYG@tok@nf{\def\PYG@tc##1{\textcolor[rgb]{0.02,0.16,0.49}{##1}}}
\def\PYG@tok@si{\let\PYG@it=\textit\def\PYG@tc##1{\textcolor[rgb]{0.44,0.63,0.82}{##1}}}
\def\PYG@tok@s2{\def\PYG@tc##1{\textcolor[rgb]{0.25,0.44,0.63}{##1}}}
\def\PYG@tok@vi{\def\PYG@tc##1{\textcolor[rgb]{0.73,0.38,0.84}{##1}}}
\def\PYG@tok@nt{\let\PYG@bf=\textbf\def\PYG@tc##1{\textcolor[rgb]{0.02,0.16,0.45}{##1}}}
\def\PYG@tok@nv{\def\PYG@tc##1{\textcolor[rgb]{0.73,0.38,0.84}{##1}}}
\def\PYG@tok@s1{\def\PYG@tc##1{\textcolor[rgb]{0.25,0.44,0.63}{##1}}}
\def\PYG@tok@gp{\let\PYG@bf=\textbf\def\PYG@tc##1{\textcolor[rgb]{0.78,0.36,0.04}{##1}}}
\def\PYG@tok@sh{\def\PYG@tc##1{\textcolor[rgb]{0.25,0.44,0.63}{##1}}}
\def\PYG@tok@ow{\let\PYG@bf=\textbf\def\PYG@tc##1{\textcolor[rgb]{0.00,0.44,0.13}{##1}}}
\def\PYG@tok@sx{\def\PYG@tc##1{\textcolor[rgb]{0.78,0.36,0.04}{##1}}}
\def\PYG@tok@bp{\def\PYG@tc##1{\textcolor[rgb]{0.00,0.44,0.13}{##1}}}
\def\PYG@tok@c1{\let\PYG@it=\textit\def\PYG@tc##1{\textcolor[rgb]{0.25,0.50,0.56}{##1}}}
\def\PYG@tok@kc{\let\PYG@bf=\textbf\def\PYG@tc##1{\textcolor[rgb]{0.00,0.44,0.13}{##1}}}
\def\PYG@tok@c{\let\PYG@it=\textit\def\PYG@tc##1{\textcolor[rgb]{0.25,0.50,0.56}{##1}}}
\def\PYG@tok@mf{\def\PYG@tc##1{\textcolor[rgb]{0.13,0.50,0.31}{##1}}}
\def\PYG@tok@err{\def\PYG@bc##1{\fcolorbox[rgb]{1.00,0.00,0.00}{1,1,1}{##1}}}
\def\PYG@tok@kd{\let\PYG@bf=\textbf\def\PYG@tc##1{\textcolor[rgb]{0.00,0.44,0.13}{##1}}}
\def\PYG@tok@ss{\def\PYG@tc##1{\textcolor[rgb]{0.32,0.47,0.09}{##1}}}
\def\PYG@tok@sr{\def\PYG@tc##1{\textcolor[rgb]{0.14,0.33,0.53}{##1}}}
\def\PYG@tok@mo{\def\PYG@tc##1{\textcolor[rgb]{0.13,0.50,0.31}{##1}}}
\def\PYG@tok@mi{\def\PYG@tc##1{\textcolor[rgb]{0.13,0.50,0.31}{##1}}}
\def\PYG@tok@kn{\let\PYG@bf=\textbf\def\PYG@tc##1{\textcolor[rgb]{0.00,0.44,0.13}{##1}}}
\def\PYG@tok@o{\def\PYG@tc##1{\textcolor[rgb]{0.40,0.40,0.40}{##1}}}
\def\PYG@tok@kr{\let\PYG@bf=\textbf\def\PYG@tc##1{\textcolor[rgb]{0.00,0.44,0.13}{##1}}}
\def\PYG@tok@s{\def\PYG@tc##1{\textcolor[rgb]{0.25,0.44,0.63}{##1}}}
\def\PYG@tok@kp{\def\PYG@tc##1{\textcolor[rgb]{0.00,0.44,0.13}{##1}}}
\def\PYG@tok@w{\def\PYG@tc##1{\textcolor[rgb]{0.73,0.73,0.73}{##1}}}
\def\PYG@tok@kt{\def\PYG@tc##1{\textcolor[rgb]{0.56,0.13,0.00}{##1}}}
\def\PYG@tok@sc{\def\PYG@tc##1{\textcolor[rgb]{0.25,0.44,0.63}{##1}}}
\def\PYG@tok@sb{\def\PYG@tc##1{\textcolor[rgb]{0.25,0.44,0.63}{##1}}}
\def\PYG@tok@k{\let\PYG@bf=\textbf\def\PYG@tc##1{\textcolor[rgb]{0.00,0.44,0.13}{##1}}}
\def\PYG@tok@se{\let\PYG@bf=\textbf\def\PYG@tc##1{\textcolor[rgb]{0.25,0.44,0.63}{##1}}}
\def\PYG@tok@sd{\let\PYG@it=\textit\def\PYG@tc##1{\textcolor[rgb]{0.25,0.44,0.63}{##1}}}

\makeatother

\begin{document}

\maketitle
\tableofcontents
\phantomsection\label{index::doc}

\section{Signal Types}
\label{signal_types:signaltypes}\label{signal_types::doc}\label{signal_types:tia-documentation-of-tobi-interface-a-version}\label{signal_types:signal-types}
TiA defines the following signals types.

\begin{tabulary}{\linewidth}{|L|L|L|L|}
\hline
\textbf{
Identifier (string)
} & \textbf{
Value (hex)
} & \textbf{
Value (dec)
} & \textbf{
Description
}\\
\hline

\code{eeg}
 & 
0x00000001
 & 
1
 & 
Electroencephalogram
\\

\code{emg}
 & 
0x00000002
 & 
2
 & 
Electromyogram
\\

\code{eog}
 & 
0x00000004
 & 
4
 & 
Electrooculogram
\\

\code{ecg}
 & 
0x00000008
 & 
8
 & 
Electrocardiogram
\\

\code{hr}
 & 
0x00000010
 & 
16
 & 
Heart rate
\\

\code{bp}
 & 
0x00000020
 & 
32
 & 
Blood pressure
\\

\code{button}
 & 
0x00000040
 & 
64
 & 
Buttons (aperiodic)
\\

\code{joystick}
 & 
0x00000080
 & 
128
 & 
Joystick axis (aperiodic)
\\

\code{sensors}
 & 
0x00000100
 & 
256
 & 
Sensor
\\

\code{nirs}
 & 
0x00000200
 & 
512
 & 
NIRS
\\

\code{fmri}
 & 
0x00000400
 & 
1,024
 & 
FMRI
\\

\code{mouse}
 & 
0x00000800
 & 
2,048
 & 
Mouse axis (aperiodic)
\\

\code{mouse-button}
 & 
0x00001000
 & 
4,096
 & 
Mouse buttons (aperiodic)
\\

not used yet
 &  &  & \\

\code{user\_1}
 & 
0x00010000
 & 
65,536
 & 
User 1
\\

\code{user\_2}
 & 
0x00020000
 & 
131,072
 & 
User 2
\\

\code{user\_3}
 & 
0x00040000
 & 
262,144
 & 
User 3
\\

\code{user\_4}
 & 
0x00080000
 & 
524,288
 & 
User 4
\\

\code{undefined}
 & 
0x00100000
 & 
1,048,576
 & 
undefined signal type
\\

\code{event}
 & 
0x00200000
 & 
2,097,152
 & 
event
\\
\hline
\end{tabulary}

\subsection{Aperiodic Signals}
\label{signal_types:aperiodic-signals}
Aperiodic signals are signals which are not transmitted at constant time rates. For example the state of joystick
axis.

Therefore values of the current state of these signals have be transmitted only if they have changed.

\section{Control Connection}
\label{control_connection:control-connection}\label{control_connection::doc}
A TiA server has to provide a TCP port on which clients can create a control connection.
Each client gets its own control connection to the server.

The control connection can be used by the clients to send requests to the server such as
``start/stop data transmission''.

Some important remarks:
\begin{enumerate}
\item {} 
The protocol is in the style of HTTP (line structured text messages)

\item {} 
The messages are encoded in UTF-8

\item {} 
The message is split into lines which are terminated by the 0x0A character (also known as \code{\textbackslash{}n}, ``line feed'' or \textless{}LF\textgreater{})

\item {} 
Some messages contain additional XML-structured content which is UTF-8 encoded

\item {} 
All characters are case sensitive!

\end{enumerate}

\subsection{Control Message Structure}
\label{control_connection:control-message-structure}
Each message which is send from the client to the server or vice versa is structured as followed:
\begin{enumerate}
\item {} 
Version line

\item {} 
Command line

\item {} 
Optional content description line

\item {} 
An empty line

\item {} 
Optional xml-structured content

\end{enumerate}

Example:

\begin{Verbatim}[commandchars=@\[\]]
TiA 1.0\n
CheckProtocolVersion\n
\n
\end{Verbatim}

Example with additional xml-structured content

\begin{Verbatim}[commandchars=@\[\]]
TiA 1.0\n
MetaInfo\n
Content-Length: 79\n
\n
@textless[]?xml version="1.0" encoding="UTF-8"?@textgreater[]@textless[]tiaMetaInfo version="1.0"@textgreater[]@textless[]/tiaMetaInfo@textgreater[]
\end{Verbatim}

\subsection{Reply Messages}
\label{control_connection:reply-messages}
Each command message is answered with an reply message which contains either an OK or an Error.
Error messages optionally contain an error description.

OK message:

\begin{Verbatim}[commandchars=@\[\]]
TiA 1.0\n
OK\n
\n
\end{Verbatim}

Error message without an error description:

\begin{Verbatim}[commandchars=@\[\]]
TiA 1.0\n
Error\n
\n
\end{Verbatim}

Error message including an error description:

\begin{Verbatim}[commandchars=@\[\]]
TiA 1.0\n
Error\n
Content-Length: 73\n
\n
@textless[]tiaError version="1.0" description="Human readable error description."/@textgreater[]
\end{Verbatim}

\subsection{Server Commands}
\label{control_connection:server-commands}
A TiA 1.0 server implementation has to support the following commands:
\begin{itemize}
\item {} 
Check protocol version

\item {} 
Get metainfo

\item {} 
Get data connection

\item {} 
Start data transmission

\item {} 
Stop data transmission

\item {} 
Get server state connection

\end{itemize}

\subsubsection{Check Protocol Version}
\label{control_connection:check-protocol-version}
This command may be used by the client to check if the server understands the commands the client wants to send.
The server has to respond with an OK message if it understands commands of the given protocol version.

Representation:

\begin{Verbatim}[commandchars=@\[\]]
TiA 1.0\n
CheckProtocolVersion\n
\n
\end{Verbatim}

Server responses either with an OK or an error message.

\subsubsection{Get MetaInfo}
\label{control_connection:get-metainfo}
This command is used to get the informations about the signals from the server.

Representation:

\begin{Verbatim}[commandchars=@\[\]]
TiA 1.0\n
GetMetaInfo\n
\n
\end{Verbatim}

Server response:

\begin{Verbatim}[commandchars=@\[\]]
TiA 1.0\n
MetaInfo\n
Content-Length: @PYGZlb[]Length of XML Content in Bytes@PYGZrb[]\n
\n
@textless[]?xml version="1.0" encoding="UTF-8"?@textgreater[]@textless[]tiaMetaInfo version="1.0"@textgreater[]....@textless[]/tiaMetaInfo@textgreater[]
\end{Verbatim}

or an error message.

\subsubsection{Get Data Transmission}
\label{control_connection:get-data-transmission}
Two types of data transmissions exist: ``TCP'' and ``UDP''.

Representation:

\begin{Verbatim}[commandchars=@\[\]]
TiA 1.0 \n
GetDataConnection: TCP \n
\n
\end{Verbatim}

or

\begin{Verbatim}[commandchars=@\[\]]
TiA 1.0 \n
GetDataConnection: UDP \n
\n
\end{Verbatim}

Server Response:

\begin{Verbatim}[commandchars=@\[\]]
TiA 1.0 \n
DataConnectionPort: @PYGZlb[]Port-Number@PYGZrb[] \n
\n
\end{Verbatim}

or an error message.

\subsubsection{Start Data Transmission}
\label{control_connection:start-data-transmission}
Representation:

\begin{Verbatim}[commandchars=@\[\]]
TiA 1.0 \n
StartDataTransmission \n
\n
\end{Verbatim}

Server responses either with an OK or an error message.

\subsubsection{Stop Data Transmission}
\label{control_connection:stop-data-transmission}
Representation:

\begin{Verbatim}[commandchars=@\[\]]
TiA 1.0 \n
StopDataTransmission \n
\n
\end{Verbatim}

Server responses either with an OK or an error message.

\subsubsection{Get Server State Connection}
\label{control_connection:get-server-state-connection}
Representation:

\begin{Verbatim}[commandchars=@\[\]]
TiA 1.0\n
GetServerStateConnection\n
\n
\end{Verbatim}

Server Response:

\begin{Verbatim}[commandchars=@\[\]]
TiA 1.0\n
ServerStateConnectionPort: @PYGZlb[]Port-Number@PYGZrb[]\n
\n
\end{Verbatim}

or an error message.

\subsection{TiA Meta Info}
\label{control_connection:tia-meta-info}
The TiA meta info is structured in XML and contains information about the signals and the subject.

\subsection{TiA Error Description}
\label{control_connection:tia-error-description}
An error message in TiA version 1.0 optionally supports an error description in a human readable format.
No error codes with special meaning are supported in this version.

\section{Server State Connection}
\label{server_state:server-state-connection}\label{server_state::doc}
Additionally to the control- and data-connection a client may request a so called ``server state connection''.

This connection is used by the server to transmit messages about its state to the clients.

\subsection{State Messages}
\label{server_state:state-messages}\begin{itemize}
\item {} 
Server running

\item {} 
Server shut down

\end{itemize}

\subsubsection{Server Running}
\label{server_state:server-running}
Just to indicate that the server is running.

\begin{Verbatim}[commandchars=@\[\]]
TiA 1.0\n
ServerStateRunning\n
\n
\end{Verbatim}

Clients must no reply to this message.

\subsubsection{Server Shut Down}
\label{server_state:server-shut-down}
Indicates that the server will shut down soon. Therefore the control and the data connection will be closed by the server soon.

\begin{Verbatim}[commandchars=@\[\]]
TiA 1.0\n
ServerStateShutdown\n
\n
\end{Verbatim}

Clients must not reply to this message.

\section{Data Packet}
\label{data_packet:data-packet}\label{data_packet::doc}
\includegraphics{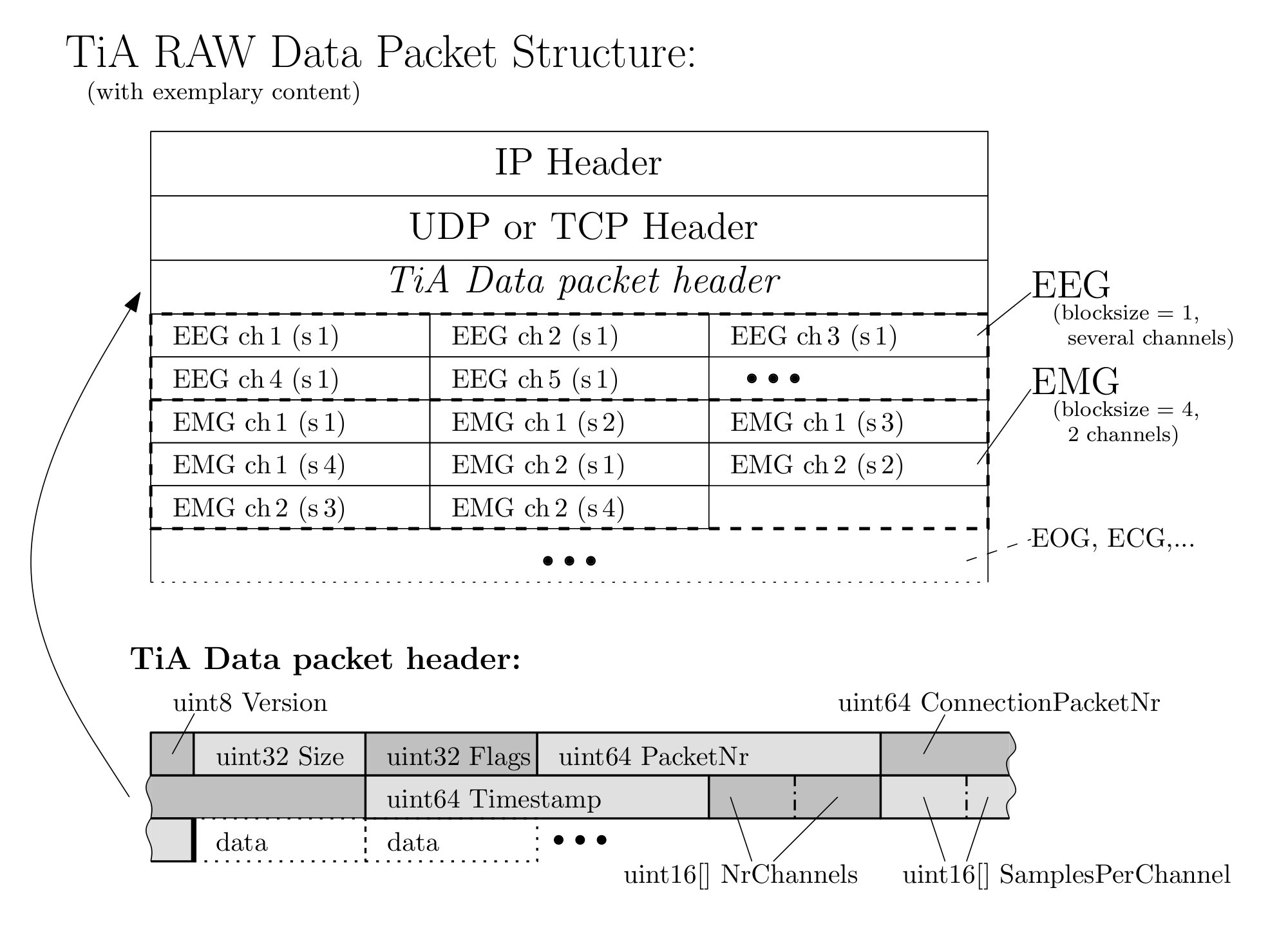}

\subsection{Fixed Header}
\label{data_packet:fixed-header}
\begin{tabulary}{\linewidth}{|L|L|L|L|}
\hline
\textbf{
Position (Byte)
} & \textbf{
Length (Byte)
} & \textbf{
Type
} & \textbf{
Content
}\\
\hline

0
 & 
1
 & 
unsigned integer
 & 
Datapacket Version has to be 3
\\

1
 & 
4
 & 
unsigned integer
 & 
Packet size (in bytes), little endian
\\

5
 & 
4
 & 
unsigned integer
 & 
Signal type flags
\\

9
 & 
8
 & 
unsigned integer
 & 
Packet id
\\

17
 & 
8
 & 
unsigned integer
 & 
Connection packet number
\\

25
 & 
8
 & 
unsigned integer
 & 
Time stamp
\\
\hline
\end{tabulary}

\subsubsection{Byte Order}
\label{data_packet:byte-order}
All data and fields are transmitted in little endian.

\subsubsection{Packet Size}
\label{data_packet:packet-size}
The packet size is the total number of bytes (octets) of a datapacket including the fixed header,
the variable header and the data.

The packet size of an empty datapacket is 33.

\subsubsection{Time Stamp}
\label{data_packet:time-stamp}
microseconds since server start

\subsection{Variable Header}
\label{data_packet:variable-header}
Remarks:
\begin{itemize}
\item {} 
NoS is the number of signals. This number can be determined by counting the flags which are set
in the ``signal type flags'' field of the fixed header.

\end{itemize}

\begin{tabulary}{\linewidth}{|L|L|L|L|}
\hline
\textbf{
Position (Byte)
} & \textbf{
Length (Byte)
} & \textbf{
Type
} & \textbf{
Content
}\\
\hline

33
 & 
2 * NoS
 & 
Array of 16bit unsigned integer
 & 
Number of channels
\\

33 + (2 * NoS)
 & 
2 * NoS
 & 
Array of 16bit unsigned integer
 & 
Block size of signal (number of samples per channel)
\\
\hline
\end{tabulary}

\subsection{Data}
\label{data_packet:data}
Data starts at the position 33 + (4 * NoS). Sample values are transmitted as a sequence of 32bit floats (IEEE 754).

Samples are firstly ordered by their signal type number ({\hyperref[signal_types:signaltypes]{\emph{Singal Types}}}).

\section{Outlook}
\label{outlook:outlook}\label{outlook::doc}
Future versions of TiA should guarantee downward compatibility. That means all TiA 1.0 commands should be supported by at least
all 1.x versions.

\subsection{Error Messages}
\label{outlook:error-messages}
Error messages may contain error codes to automatically interpret the meaning of an error.

\subsection{Datapackets}
\label{outlook:datapackets}
Important remark: Due to downward compatibility no datapacket can become version 10! (as the first byte
in datapacket version 2 is ``10'').

\section{Appendix}
\label{appendix:appendix}\label{appendix::doc}

\subsection{TiA MetaInfo XML Schema}
\label{appendix:tia-metainfo-xml-schema}
\begin{Verbatim}[commandchars=\\\{\},numbers=left,firstnumber=1,stepnumber=1]
\PYG{c+cp}{\textless{}?xml version="1.0" encoding="UTF-8"?\textgreater{}}
\PYG{c}{\textless{}!--}
\PYG{c}{  This is the XML Schema for the meta info XML representation version 1.0}
\PYG{c}{  of TOBI Interface A (TiA)}
\PYG{c}{  }
\PYG{c}{  Version 1.0}
\PYG{c}{  }
\PYG{c}{  author: Christoph Eibel}
\PYG{c}{--\textgreater{}}
\PYG{n+nt}{\textless{}xsd:schema} \PYG{n+na}{xmlns:xsd=}\PYG{l+s}{"http://www.w3.org/2001/XMLSchema"}\PYG{n+nt}{\textgreater{}}
  \PYG{n+nt}{\textless{}xsd:element} \PYG{n+na}{name=}\PYG{l+s}{"tiaMetaInfo"}\PYG{n+nt}{\textgreater{}}
    \PYG{n+nt}{\textless{}xsd:complexType}\PYG{n+nt}{\textgreater{}}
      \PYG{n+nt}{\textless{}xsd:sequence}\PYG{n+nt}{\textgreater{}}
	
	\PYG{c}{\textless{}!--}\PYG{c}{ subject info }\PYG{c}{--\textgreater{}}
	\PYG{n+nt}{\textless{}xsd:element} \PYG{n+na}{name=}\PYG{l+s}{"subject"} \PYG{n+na}{minOccurs=}\PYG{l+s}{"0"} \PYG{n+na}{maxOccurs=}\PYG{l+s}{"1"}\PYG{n+nt}{\textgreater{}}
	  \PYG{n+nt}{\textless{}xsd:complexType}\PYG{n+nt}{\textgreater{}}
	    \PYG{n+nt}{\textless{}xsd:attribute} \PYG{n+na}{name=}\PYG{l+s}{"id"} \PYG{n+na}{type=}\PYG{l+s}{"xsd:string"} \PYG{n+na}{use=}\PYG{l+s}{"optional"}\PYG{n+nt}{/\textgreater{}}
	    \PYG{n+nt}{\textless{}xsd:attribute} \PYG{n+na}{name=}\PYG{l+s}{"firstName"} \PYG{n+na}{type=}\PYG{l+s}{"xsd:string"} \PYG{n+na}{use=}\PYG{l+s}{"optional"}\PYG{n+nt}{/\textgreater{}}
	    \PYG{n+nt}{\textless{}xsd:attribute} \PYG{n+na}{name=}\PYG{l+s}{"surname"} \PYG{n+na}{type=}\PYG{l+s}{"xsd:string"} \PYG{n+na}{use=}\PYG{l+s}{"optional"}\PYG{n+nt}{/\textgreater{}}
	    \PYG{n+nt}{\textless{}xsd:attribute} \PYG{n+na}{name=}\PYG{l+s}{"sex"} \PYG{n+na}{use=}\PYG{l+s}{"optional"}\PYG{n+nt}{\textgreater{}}
	      \PYG{n+nt}{\textless{}xsd:simpleType}\PYG{n+nt}{\textgreater{}}
		\PYG{n+nt}{\textless{}xsd:restriction} \PYG{n+na}{base=}\PYG{l+s}{"xsd:string"}\PYG{n+nt}{\textgreater{}}
		  \PYG{n+nt}{\textless{}xsd:enumeration} \PYG{n+na}{value=}\PYG{l+s}{"m"} \PYG{n+nt}{/\textgreater{}}
		  \PYG{n+nt}{\textless{}xsd:enumeration} \PYG{n+na}{value=}\PYG{l+s}{"f"} \PYG{n+nt}{/\textgreater{}}
		\PYG{n+nt}{\textless{}/xsd:restriction\textgreater{}}
	      \PYG{n+nt}{\textless{}/xsd:simpleType\textgreater{}}
	    \PYG{n+nt}{\textless{}/xsd:attribute\textgreater{}}
	    \PYG{n+nt}{\textless{}xsd:attribute} \PYG{n+na}{name=}\PYG{l+s}{"birthday"} \PYG{n+na}{type=}\PYG{l+s}{"xsd:date"} \PYG{n+na}{use=}\PYG{l+s}{"optional"}\PYG{n+nt}{/\textgreater{}}
	    \PYG{n+nt}{\textless{}xsd:attribute} \PYG{n+na}{name=}\PYG{l+s}{"handedness"} \PYG{n+na}{use=}\PYG{l+s}{"optional"}\PYG{n+nt}{\textgreater{}}
	      \PYG{n+nt}{\textless{}xsd:simpleType}\PYG{n+nt}{\textgreater{}}
		\PYG{n+nt}{\textless{}xsd:restriction} \PYG{n+na}{base=}\PYG{l+s}{"xsd:string"}\PYG{n+nt}{\textgreater{}}
		  \PYG{n+nt}{\textless{}xsd:enumeration} \PYG{n+na}{value=}\PYG{l+s}{"l"} \PYG{n+nt}{/\textgreater{}}
		  \PYG{n+nt}{\textless{}xsd:enumeration} \PYG{n+na}{value=}\PYG{l+s}{"r"} \PYG{n+nt}{/\textgreater{}}
		\PYG{n+nt}{\textless{}/xsd:restriction\textgreater{}}
	      \PYG{n+nt}{\textless{}/xsd:simpleType\textgreater{}}	      
	    \PYG{n+nt}{\textless{}/xsd:attribute\textgreater{}}
	    \PYG{n+nt}{\textless{}xsd:attribute} \PYG{n+na}{name=}\PYG{l+s}{"medication"} \PYG{n+na}{type=}\PYG{l+s}{"xsd:boolean"} \PYG{n+na}{use=}\PYG{l+s}{"optional"} \PYG{n+nt}{/\textgreater{}}
	    \PYG{n+nt}{\textless{}xsd:attribute} \PYG{n+na}{name=}\PYG{l+s}{"glasses"} \PYG{n+na}{type=}\PYG{l+s}{"xsd:boolean"} \PYG{n+na}{use=}\PYG{l+s}{"optional"} \PYG{n+nt}{/\textgreater{}}
	    \PYG{n+nt}{\textless{}xsd:attribute} \PYG{n+na}{name=}\PYG{l+s}{"smoker"} \PYG{n+na}{type=}\PYG{l+s}{"xsd:boolean"} \PYG{n+na}{use=}\PYG{l+s}{"optional"} \PYG{n+nt}{/\textgreater{}}
	  \PYG{n+nt}{\textless{}/xsd:complexType\textgreater{}}
	\PYG{n+nt}{\textless{}/xsd:element\textgreater{}}
	
	\PYG{c}{\textless{}!--}\PYG{c}{ master signal info }\PYG{c}{--\textgreater{}}
	\PYG{n+nt}{\textless{}xsd:element} \PYG{n+na}{name=}\PYG{l+s}{"masterSignal"} \PYG{n+na}{minOccurs=}\PYG{l+s}{"0"} \PYG{n+na}{maxOccurs=}\PYG{l+s}{"1"}\PYG{n+nt}{\textgreater{}}
	  \PYG{n+nt}{\textless{}xsd:complexType}\PYG{n+nt}{\textgreater{}}
	    \PYG{n+nt}{\textless{}xsd:attribute} \PYG{n+na}{name=}\PYG{l+s}{"samplingRate"} \PYG{n+na}{type=}\PYG{l+s}{"xsd:float"} \PYG{n+na}{use=}\PYG{l+s}{"required"}\PYG{n+nt}{/\textgreater{}}
	    \PYG{n+nt}{\textless{}xsd:attribute} \PYG{n+na}{name=}\PYG{l+s}{"blockSize"} \PYG{n+na}{type=}\PYG{l+s}{"xsd:integer"} \PYG{n+na}{use=}\PYG{l+s}{"required"}\PYG{n+nt}{/\textgreater{}}
	  \PYG{n+nt}{\textless{}/xsd:complexType\textgreater{}}
	\PYG{n+nt}{\textless{}/xsd:element\textgreater{}}
	  
	
	\PYG{c}{\textless{}!--}\PYG{c}{ signal info }\PYG{c}{--\textgreater{}}
	\PYG{n+nt}{\textless{}xsd:element} \PYG{n+na}{name=}\PYG{l+s}{"signal"} \PYG{n+na}{minOccurs=}\PYG{l+s}{"0"} \PYG{n+na}{maxOccurs=}\PYG{l+s}{"unbounded"}\PYG{n+nt}{\textgreater{}}
	  \PYG{n+nt}{\textless{}xsd:complexType}\PYG{n+nt}{\textgreater{}}
	    
	    \PYG{c}{\textless{}!--}\PYG{c}{ channels }\PYG{c}{--\textgreater{}}
	    \PYG{n+nt}{\textless{}xsd:sequence}\PYG{n+nt}{\textgreater{}}
	      \PYG{n+nt}{\textless{}xsd:element} \PYG{n+na}{name=}\PYG{l+s}{"channel"} \PYG{n+na}{minOccurs=}\PYG{l+s}{"0"} \PYG{n+na}{maxOccurs=}\PYG{l+s}{"unbounded"}\PYG{n+nt}{\textgreater{}}
		\PYG{n+nt}{\textless{}xsd:complexType}\PYG{n+nt}{\textgreater{}}
		  \PYG{n+nt}{\textless{}xsd:attribute} \PYG{n+na}{name=}\PYG{l+s}{"nr"} \PYG{n+na}{type=}\PYG{l+s}{"xsd:nonNegativeInteger"} \PYG{n+na}{use=}\PYG{l+s}{"required"}\PYG{n+nt}{/\textgreater{}}
		  \PYG{n+nt}{\textless{}xsd:attribute} \PYG{n+na}{name=}\PYG{l+s}{"label"} \PYG{n+na}{type=}\PYG{l+s}{"xsd:string"} \PYG{n+na}{use=}\PYG{l+s}{"required"}\PYG{n+nt}{/\textgreater{}}
		\PYG{n+nt}{\textless{}/xsd:complexType\textgreater{}}
	      \PYG{n+nt}{\textless{}/xsd:element\textgreater{}}
	    \PYG{n+nt}{\textless{}/xsd:sequenc\textgreater{}}
	    
	    \PYG{c}{\textless{}!--}\PYG{c}{ signal type: as defined in TiA 1.0 }\PYG{c}{--\textgreater{}}
	    \PYG{n+nt}{\textless{}xsd:attribute} \PYG{n+na}{name=}\PYG{l+s}{"type"} \PYG{n+na}{use=}\PYG{l+s}{"required"}\PYG{n+nt}{\textgreater{}}
	      \PYG{n+nt}{\textless{}xsd:simpleType}\PYG{n+nt}{\textgreater{}}
		\PYG{n+nt}{\textless{}xsd:restriction} \PYG{n+na}{base=}\PYG{l+s}{"xsd:string"}\PYG{n+nt}{\textgreater{}}
		  \PYG{n+nt}{\textless{}xsd:enumeration} \PYG{n+na}{value=}\PYG{l+s}{"eeg"} \PYG{n+nt}{/\textgreater{}}
		  \PYG{n+nt}{\textless{}xsd:enumeration} \PYG{n+na}{value=}\PYG{l+s}{"emg"} \PYG{n+nt}{/\textgreater{}}
		  \PYG{n+nt}{\textless{}xsd:enumeration} \PYG{n+na}{value=}\PYG{l+s}{"eog"} \PYG{n+nt}{/\textgreater{}}
		  \PYG{n+nt}{\textless{}xsd:enumeration} \PYG{n+na}{value=}\PYG{l+s}{"ecg"} \PYG{n+nt}{/\textgreater{}}
		  \PYG{n+nt}{\textless{}xsd:enumeration} \PYG{n+na}{value=}\PYG{l+s}{"hr"} \PYG{n+nt}{/\textgreater{}}
		  \PYG{n+nt}{\textless{}xsd:enumeration} \PYG{n+na}{value=}\PYG{l+s}{"bp"} \PYG{n+nt}{/\textgreater{}}
		  \PYG{n+nt}{\textless{}xsd:enumeration} \PYG{n+na}{value=}\PYG{l+s}{"buttons"} \PYG{n+nt}{/\textgreater{}}
		  \PYG{n+nt}{\textless{}xsd:enumeration} \PYG{n+na}{value=}\PYG{l+s}{"joystick"} \PYG{n+nt}{/\textgreater{}}
		  \PYG{n+nt}{\textless{}xsd:enumeration} \PYG{n+na}{value=}\PYG{l+s}{"sensors"} \PYG{n+nt}{/\textgreater{}}
		  \PYG{n+nt}{\textless{}xsd:enumeration} \PYG{n+na}{value=}\PYG{l+s}{"nirs"} \PYG{n+nt}{/\textgreater{}}
		  \PYG{n+nt}{\textless{}xsd:enumeration} \PYG{n+na}{value=}\PYG{l+s}{"fmri"} \PYG{n+nt}{/\textgreater{}}
		  \PYG{n+nt}{\textless{}xsd:enumeration} \PYG{n+na}{value=}\PYG{l+s}{"undefined"} \PYG{n+nt}{/\textgreater{}}
		\PYG{n+nt}{\textless{}/xsd:restriction\textgreater{}}
	      \PYG{n+nt}{\textless{}/xsd:simpleType\textgreater{}}	      
	    \PYG{n+nt}{\textless{}/xsd:attribute\textgreater{}}

	    \PYG{n+nt}{\textless{}xsd:attribute} \PYG{n+na}{name=}\PYG{l+s}{"samplingRate"} \PYG{n+na}{type=}\PYG{l+s}{"xsd:float"} \PYG{n+na}{use=}\PYG{l+s}{"required"}\PYG{n+nt}{/\textgreater{}}
	    \PYG{n+nt}{\textless{}xsd:attribute} \PYG{n+na}{name=}\PYG{l+s}{"blockSize"} \PYG{n+na}{type=}\PYG{l+s}{"xsd:integer"} \PYG{n+na}{use=}\PYG{l+s}{"required"}\PYG{n+nt}{/\textgreater{}}
	    \PYG{n+nt}{\textless{}xsd:attribute} \PYG{n+na}{name=}\PYG{l+s}{"numChannels"} \PYG{n+na}{type=}\PYG{l+s}{"xsd:nonNegativeInteger"} \PYG{n+na}{use=}\PYG{l+s}{"required"}\PYG{n+nt}{/\textgreater{}}
	  \PYG{n+nt}{\textless{}/xsd:complexType\textgreater{}}
	\PYG{n+nt}{\textless{}/xsd:element\textgreater{}}

      \PYG{n+nt}{\textless{}/xsd:sequence\textgreater{}}

      \PYG{c}{\textless{}!--}\PYG{c}{ version attribute }\PYG{c}{--\textgreater{}}
      \PYG{n+nt}{\textless{}xsd:attribute} \PYG{n+na}{name=}\PYG{l+s}{"version"} \PYG{n+na}{type=}\PYG{l+s}{"xsd:string"} \PYG{n+na}{fixed=}\PYG{l+s}{"1.0"} \PYG{n+na}{use=}\PYG{l+s}{"required"} \PYG{n+nt}{/\textgreater{}}
      
    \PYG{n+nt}{\textless{}/xsd:complexType\textgreater{}}
  \PYG{n+nt}{\textless{}/xsd:element\textgreater{}}

\PYG{n+nt}{\textless{}/xsd:schema\textgreater{}}
\end{Verbatim}

\subsection{TiA MetaInfo XML Example}
\label{appendix:tia-metainfo-xml-example}
\begin{Verbatim}[commandchars=\\\{\},numbers=left,firstnumber=1,stepnumber=1]
\PYG{c+cp}{\textless{}?xml version="1.0" encoding="UTF-8"?\textgreater{}}
\PYG{n+nt}{\textless{}tiaMetaInfo} \PYG{n+na}{version=}\PYG{l+s}{"1.0"}\PYG{n+nt}{\textgreater{}}
  
  \PYG{n+nt}{\textless{}subject} \PYG{n+na}{id=}\PYG{l+s}{"WE2"} \PYG{n+na}{firstName=}\PYG{l+s}{"Max"} \PYG{n+na}{lastName=}\PYG{l+s}{"Mustermann"} 
           \PYG{n+na}{handedness=}\PYG{l+s}{"r"} \PYG{n+nt}{/\textgreater{}}

  \PYG{n+nt}{\textless{}masterSignal} \PYG{n+na}{sampleRate=}\PYG{l+s}{"100"} \PYG{n+na}{blockSize=}\PYG{l+s}{"10"} \PYG{n+nt}{/\textgreater{}}
  
  \PYG{n+nt}{\textless{}signal} \PYG{n+na}{type=}\PYG{l+s}{"eeg"} \PYG{n+na}{blockSize=}\PYG{l+s}{"10"} \PYG{n+na}{sampleRate=}\PYG{l+s}{"100"} \PYG{n+na}{numChannels=}\PYG{l+s}{"3"}\PYG{n+nt}{\textgreater{}}
    \PYG{n+nt}{\textless{}channel} \PYG{n+na}{nr=}\PYG{l+s}{"1"} \PYG{n+na}{label=}\PYG{l+s}{"Cz"}\PYG{n+nt}{/\textgreater{}}
    \PYG{n+nt}{\textless{}channel} \PYG{n+na}{nr=}\PYG{l+s}{"2"} \PYG{n+na}{label=}\PYG{l+s}{"C1"}\PYG{n+nt}{/\textgreater{}}
    \PYG{n+nt}{\textless{}channel} \PYG{n+na}{nr=}\PYG{l+s}{"3"} \PYG{n+na}{label=}\PYG{l+s}{"C2"}\PYG{n+nt}{/\textgreater{}}
  \PYG{n+nt}{\textless{}/signal\textgreater{}}

  \PYG{n+nt}{\textless{}signal} \PYG{n+na}{type=}\PYG{l+s}{"bp"} \PYG{n+na}{blockSize=}\PYG{l+s}{"5"} \PYG{n+na}{sampleRate=}\PYG{l+s}{"50"} \PYG{n+na}{numChannels=}\PYG{l+s}{"5"}\PYG{n+nt}{\textgreater{}}
    \PYG{n+nt}{\textless{}channel} \PYG{n+na}{nr=}\PYG{l+s}{"3"} \PYG{n+na}{label=}\PYG{l+s}{"Channel 3 with Label"}\PYG{n+nt}{/\textgreater{}}
    \PYG{n+nt}{\textless{}channel} \PYG{n+na}{nr=}\PYG{l+s}{"2"} \PYG{n+na}{label=}\PYG{l+s}{"Channel 2 with Label"}\PYG{n+nt}{/\textgreater{}}
    \PYG{c}{\textless{}!--}
\PYG{c}{      Channel 1, 4 and 5 have no labels}
\PYG{c}{      }\PYG{c}{--\textgreater{}}
  \PYG{n+nt}{\textless{}/signal\textgreater{}}
  
\PYG{n+nt}{\textless{}/tiaMetaInfo\textgreater{}}
\end{Verbatim}

\subsection{TiA Error Message XML Schema}
\label{appendix:tia-error-message-xml-schema}

\renewcommand{\indexname}{Index}
\printindex
\end{document}